\documentclass[prb,preprint]{revtex4-1} 

\usepackage{amsmath}  
\usepackage{amsfonts} 
\usepackage{graphicx} 
\usepackage{subcaption} 

\graphicspath{{Images/}}

\newcommand{\Equation}[1]{\text{Eq.~}\eqref{#1}}
\newcommand{\Fig}[1]{\text{Fig.~}\ref{#1}}

\begin{document}


\title{The Solar Solution: Tracking the Sun with Low-Energy Neutrinos}

\author{Nicole Hartman}
\author{Stephen Sekula}
\affiliation{Department of Physics, Southern Methodist University, Dallas, TX 75275-0175}



\date{\today}

\begin{abstract}

As neutrinos become a significant background for projected dark matter experiments, the community will become concerned with determining if events counted in a dark matter experiment are good dark matter candidates or low-energy neutrinos from astrophysical sources.  We investigate the feasibility of using neutrino-electron scattering in a terrestrial detector medium as a means to determine the flight direction of the original, low-energy solar neutrino.
Using leading-order weak interactions in the Standard Model and constrains from energy and momentum conservation, we developed a simple simulation that suggests that~68\% of the time the ejected electron would be within 0.99 radians of the incident neutrino's direction.  This suggests that it may be fruitful to pursue low-energy neutrino detection capability that can utilize such ejected electrons.

\end{abstract}

\maketitle 

\section{Introduction} 

\begin{figure}[h!tbp]
\centering
\includegraphics[scale=0.50]{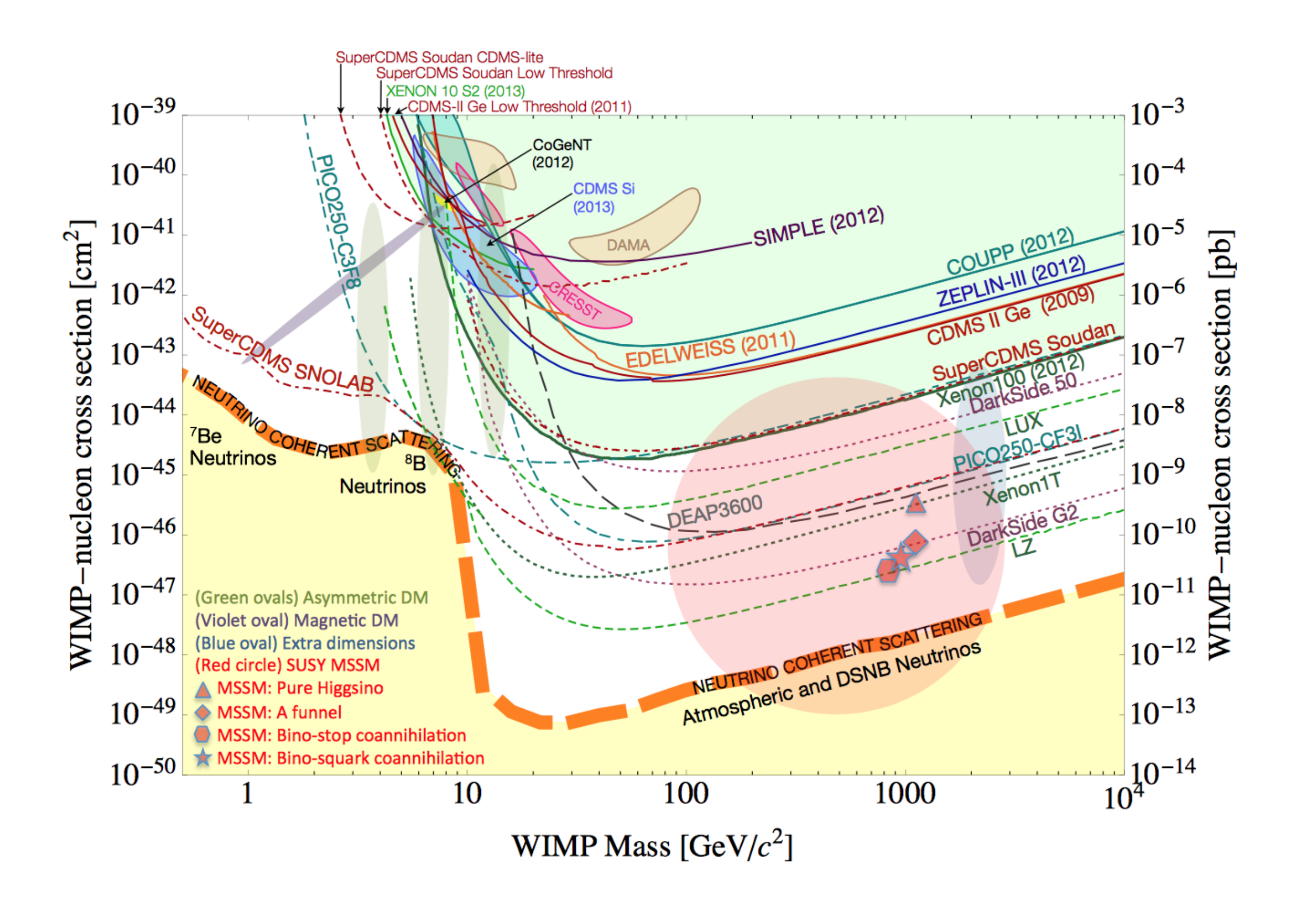}
\caption{The cross-section limits (solid curves) for the WIMP-nucleon spin independent interaction and projection (dashed curves) for future direct detection experiments expected in the next decade.  Of particular interest to this analysis, the lowest sweeping dashed orange band indicates the projected sensitively for WIMP experiments to backgrounds of solar, atmospheric and diffuse supernovae neutrinos.\cite{Feng:2014uja}}
\label{Fig:sensitivities}
\end{figure}

The nature of dark matter nature is one of the major scientific questions the modern era.  Approximately 86\% of the matter of the universe is not luminous, and observed only by its gravitational interaction.\cite{Frieman:2008,Frenk:2012}  The neutral, weakly interacting neutrino from the Standard Model was once considered a dark matter candidate, but further measurements proved that these particles cannot account for the the entirety of the dark matter.\cite{Frenk:2012,White:1983,Michael:2006}  Although neutrinos were no longer a serious candidate for dark matter experiments, they could still be a background.\cite{Feng:2014uja}  Early dark matter experiments did not account for the neutrino flux because they were not sensitive to it.  However, as dark matter remains undetected and detector sensitivity has increased, the relevant backgrounds will need to be precisely measured and subtracted.\cite{Billard:2014}  In \Fig{Fig:sensitivities} we show the projected sensitivities for future dark matter experiments.\cite{Feng:2014uja}  The thick, dashed line on this graph indicates at what cross-sections for various candidate WIMP masses the experiments will be sensitive to neutrino backgrounds.  

\begin{figure}[h!tbp]
\centering
\includegraphics[width=0.8\textwidth]{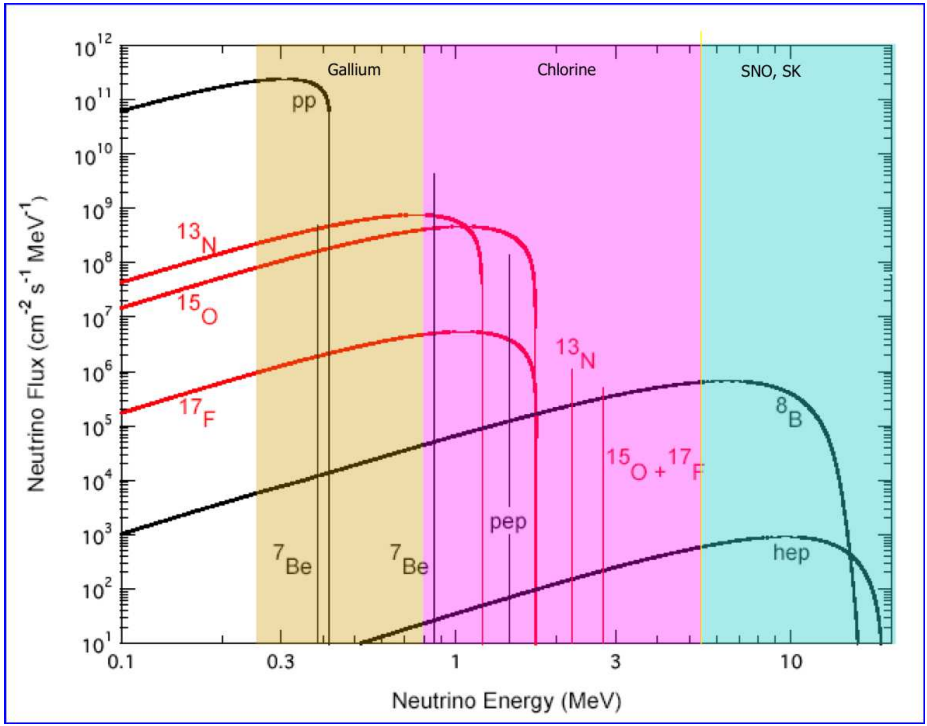}
\caption{Energy dependence for the various neutrino sources with the detectors that are sensitive to each of these signals.~\cite{Chieze:2010}}
\label{Fig:neutrinoFlux}
\end{figure}

Since the sun is the closest cosmic nuclear reactor, it should produce the largest flux of neutrinos close to earth.
There are many different reactions that the sun undergoes as it produces photons, but the most common one is known as the ``pp chain.''\cite{Moaz, Chieze:2010}  In \Fig{Fig:neutrinoFlux} we show the fluxes as a function of the neutrino energies produced by other nuclear reactions in the sun.\cite{Chieze:2010}

The first step in the sequence of reactions for the pp chain is
\begin{equation}
p + p \rightarrow d + e^+ + \nu_e,
\label{Eq:pp_chain}
\end{equation}
which gives off 0.420~MeV as a proton turns into a neutron to form deuteron.  The proton, deuteron, and electron have rest masses of 938.27~MeV, 1875.6~MeV, and 0.511~MeV, respectively, so the maximum amount of energy that a massless neutrino could carry away would be
\begin{equation}
\text{max}(E_\nu) = [2(938.27)-1875.6-0.511] \text{MeV} = 0.420~\text{MeV}.
\end{equation}

The neutrino's flight direction could help determine whether a given dark matter candidate event could be background.  If solar neutrinos were to enter an active detector volume, a fraction of these neutrinos could interact with the atoms to eject valence electrons.  Assuming technology will be developed to allow for tracking such electrons, we wanted to determine the feasibility of correlating the direction of the ejected electron with the flight direction of the incident neutrino.  This simulation that we developed predicts this correlation by finding the possible angles of deflection for the ejected electron.  Experimentally, smaller angles would be preferred since this better constrains the neutrino's path and therefore yields the most information about the incident solar neutrino flux.

\section{Theoretical Tools}

\begin{figure}[h!tbp]
	\centering
        \includegraphics[width=0.8\textwidth]{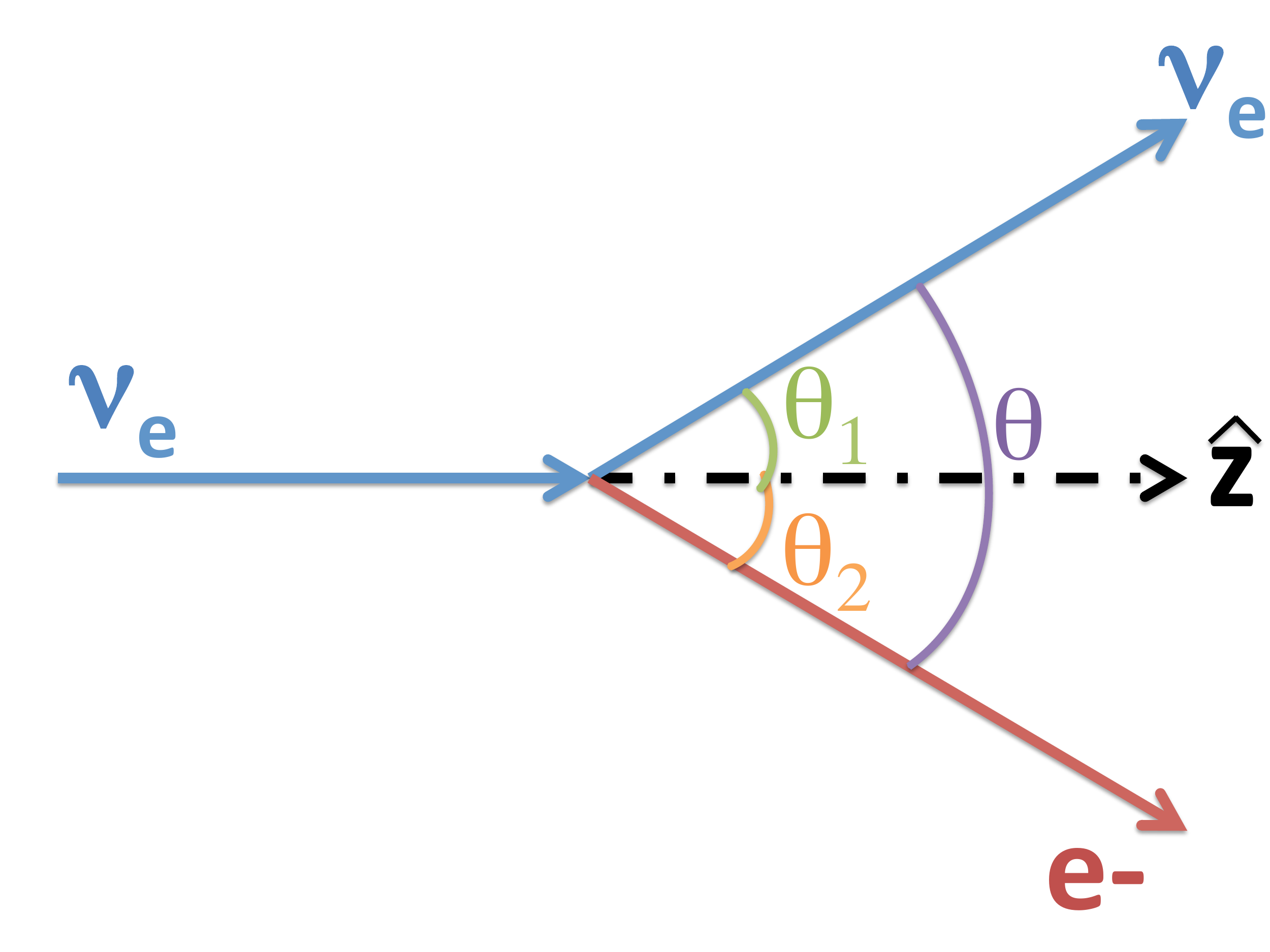}
        \caption{Schematic defining the variables for the model for the interaction.}
        \label{Fig:model}
\end{figure}

The model we used to find the angle of the ejected electron assumes that the valence electron is initially at rest.  To justify these assumptions, recall that most noble elements have a typical ionization energy of 20 eV, and by the virial theorem, this is also the kinetic energy of the electron.  The MeV order energies of the incident neutrinos are sufficiently large compared to the kinetic energies of the valence electrons, verifying the assumption of an electron at rest.  Similarly, in the model we used a massless neutrino because the neutrino mass is of the order of an eV or smaller, at least millionth of the energy of the incident neutrino.   We illustrate the relevant variables in this model in \Fig{Fig:model}. 

Conservation of energy requires
\begin{equation}
E_\nu + m_e = E_\nu' + E_e',
\label{Eq: energy conservation}
\end{equation}
where the subscript $\nu$ refers to the neutrino while the subscript $e$ pertains to the electron.  The unprimed variables correspond to the values before the collision and the primed variables denote post-collision values.  $E$ stands for energy and $m$ is for mass, because in natural units, $c = 1$ so energy and mass have the same units.

Also, the application of conservation of momentum along the flight direction of the incoming neutrino and in the transverse direction yields \Equation{Eq:p_p} and \Equation{Eq:p_t}, respectively,
\begin{equation}
p_\nu = p_\nu' \cos(\theta_1)+ p_e' \cos(\theta_2),
\label{Eq:p_p}
\end{equation}
\begin{equation}
0 = p_v' \sin(\theta_1)+ p_e' \sin(\theta_2),
\label{Eq:p_t}
\end{equation}
where $p$ stands for momentum and the same convention for subscripts and superscripts is observed as in \Equation{Eq: energy conservation}.  The variables $\theta_1$ and $\theta_2$ are defined in \Fig{Fig:model}.

Then we can relate the energies and momentums, utilizing Einstein's relation 
\begin{equation}
E_i^2  = p_i^2 + m_i^2.
\end{equation}
Since the neutrino is approximated as massless, its energy will equal its momentum.  The initial energy of the neutrino, $E_\nu$, was set to 0.420~MeV because this is the maximum energy that is given off by this reaction in the pp chain.  Therefore, applying Einstein's relation yields three equations for four unknowns: $\theta_1, \theta_2, E_\nu', \text{ and } E_e'$.  The conservation of energy and conservation of momentum equations do not admit a unique solution because there are many possible values of, for instance, the angles that can satisfy the existing constraints.  To select the most probable configurations, we employ the scattering cross-section for this process, which is proportional to the probability that such an interaction will occur.  The generic equation for the cross-section of the interaction is \Equation{cs},\cite{Hosen}
\begin{equation}
\frac{d\sigma}{dy} = \frac{G_F^2 s}{4 \pi} [ (c_V + c_A)^2 + (c_V - c_A)^2 (1 - y)^2 ],
\label{cs}
\end{equation}
where $\sigma$ is the total cross-section; $s$ is the is the square of the energy, $(E_\nu+m_e)^2$; $c_V$ and $c_A$ are the vectorial and axial couplings, respectively; and $y = \frac{1-cos(\theta)}{2}$ incorporates the angular dependence of the cross section, since $\theta \equiv \theta_1 - \theta_2$ by \Fig{Fig:model} .

\begin{figure}
\centering
\begin{subfigure}[h!btp]{0.4\textwidth}
	\includegraphics[width=\textwidth]{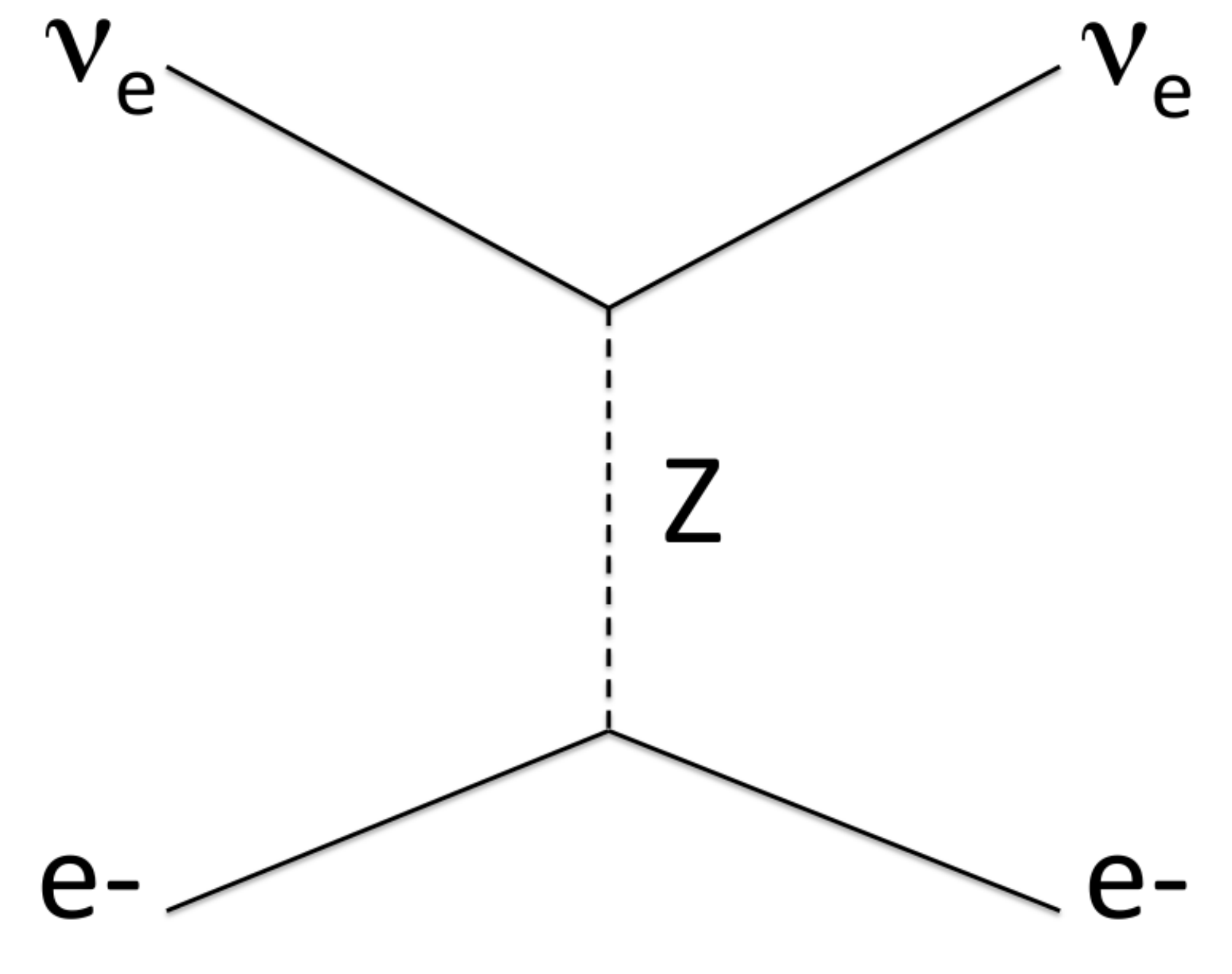}
	\caption{Interaction mediated by Z boson.}
	\label{Fig:Zcurrent}
\end{subfigure}
~
\begin{subfigure}[h!btp]{0.4\textwidth}
	\includegraphics[width=\textwidth]{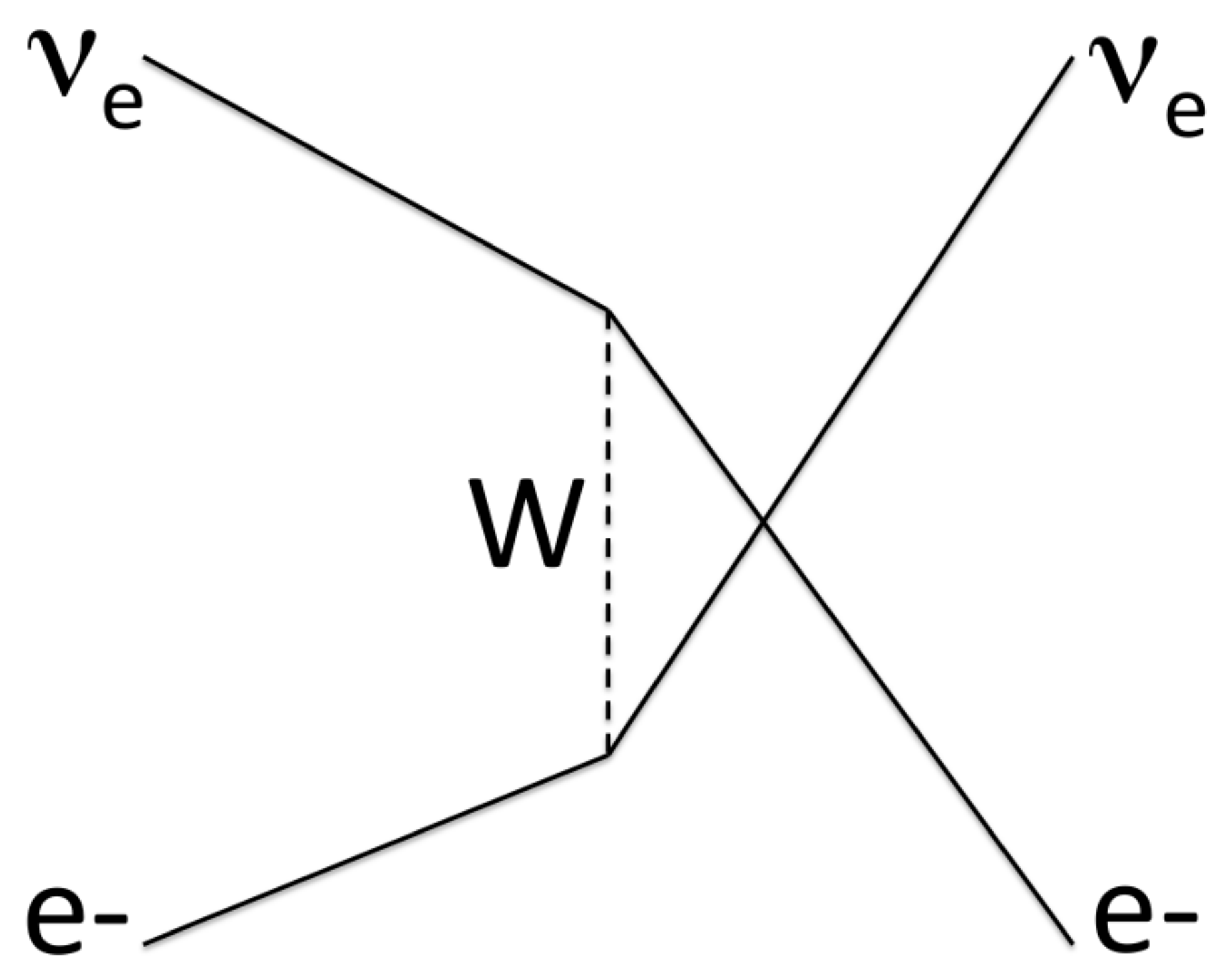}
	\caption{Interaction mediated by W boson.}
	\label{Fig:Wcurrent}
\end{subfigure}
\caption{Feynman diagrams that interfere with each other to determine the probability for for the cross-section of the interaction $\nu_e + e- \rightarrow \nu_e + e-$, given in \Equation{cs}.\cite{Hosen}}
\label{Fig:Feynman}
\end{figure}

However, the formula above is not complete, because this interaction is governed by the weak force, and it can be mediated by either the Z boson or the charged W boson.  The Feynman diagrams for these two interactions are shown in \Fig{Fig:Feynman}.
Since these two diagrams can interfere with each other, an extra interference term $\frac{G_F^2 m_e^2 y}{2 \pi}[(c_V + 1)^2 - (c_A+1)^2]$ is added to \Equation{cs}, where $G_F$ is the Fermi coupling.\cite{Hosen}  In addition, the couplings in \Equation{cs} are replaced by $c_V \rightarrow c_V + 1$ and $c_A \rightarrow c_A + 1$.  This means that the cross-section for the interaction, written out completely at leading order, is
\begin{equation}
\frac{d\sigma}{dy} = \frac{G_F^2 s}{4 \pi} [ (c_V + c_A + 2)^2 + (c_V - c_A)^2 (1 - y)^2 ] + \frac{G_F^2 m_e^2 y}{2 \pi}[(c_V + 1)^2 - (c_A+1)^2].
\label{cs_final}
\end{equation}

\section{Monte Carlo Simulation}

We can write the conservation of energy and momentum equations in terms of any one of the unknowns in the problem: $E_v', E_e', \theta_1$, or $\theta_2$.  Since the system is underdetermined, we can use the cross-section in \Equation{cs_final} as a probability distribution to find the range of allowed values for the parameter of interest.  
We use a Monte-Carlo program to run an accept-reject method to accomplish this task.
We reduced the conservation of energy and momentum equations to a single equation in terms of $E_\nu'$, and then expressed the other relevant variables in terms of $E'_\nu$.  
Since the maximum neutrino energy for solar neutrinos that can come from the pp reaction is 0.420~MeV, a Python-implemented uniform random number generator produced values for $E_\nu'$ between 0 and 0.420~MeV because all physical solutions had to lie in this range.  
However, the neutrino can never have its energy go all the way to zero because in the massless neutrino model, the neutrino must travel at the speed of light, and hence has a non-zero energy $hc / \lambda$, where $\lambda$ is the neutrino's wavelength.  
The approximation limits the energy of the neutrino to lie above a minimum that is not zero.  
We discarded values of $E_\nu'$ that produced unphysical solutions.
If the random $E_\nu'$ values allowed physical solutions for $E_e'$, $\sin(\theta_1)$, and $\sin(\theta_2)$, then the probability for the event was calculated.

The calculation was performed in two steps. 
First, the event generator found $10^6$ physical events and returned the maximum cross-section, $p_{max}$, to use in the accept-reject cycle.
The goal of the accept-reject method is to use the probability distribution to find the distribution of $\theta_2$ values.  
To implement, as the event generator looped through the events and found the cross-section, $p_{cal}$, for each physical event, another uniform random number generator generated a value, $p_{ex}$, between 0 and $p_{max}$. 
If $p_{ex} > p_{cal}$, we kept the values of the angles and particle energies used to determine $p_{cal}$.  However, if $p_{ex} < p_{cal}$, the event was rejected.  After repeating this for $10^6$ iterations, we plotted the histogram for the $\theta_2$ values, representing the scattering of $10^6$ neutrinos off atomic electrons in some detector medium.  
Then we used this histogram to find the upper limit angle for $\theta_2$ that included 68.3\%, 90\%, 95\%, and 99\% of the data.

Although 0.420~MeV neutrinos are interesting because they are the highest energy that neutrinos from the pp chain reaction can have and thus will be the easiest of the pp-neutrinos to detect, other energies were also of interest.  For example, most of the pp-neutrinos would have an energy of around 0.260~MeV,\cite{Moaz} and neutrinos from other nuclear reactions in the sun could have even larger energies, as shown in \Fig{Fig:neutrinoFlux}.  To compare energies with the values seen in \Fig{Fig:neutrinoFlux}, we made $\theta_2$ histograms for a range of incident energies.  We then determined the upper $\theta_2$ limit for different confidence levels for each of these histograms to plot the upper $\theta_2$ limit as a function of energy.  

\section{Results}

\begin{figure}[h!tbp]
\centering
\includegraphics[scale=0.50]{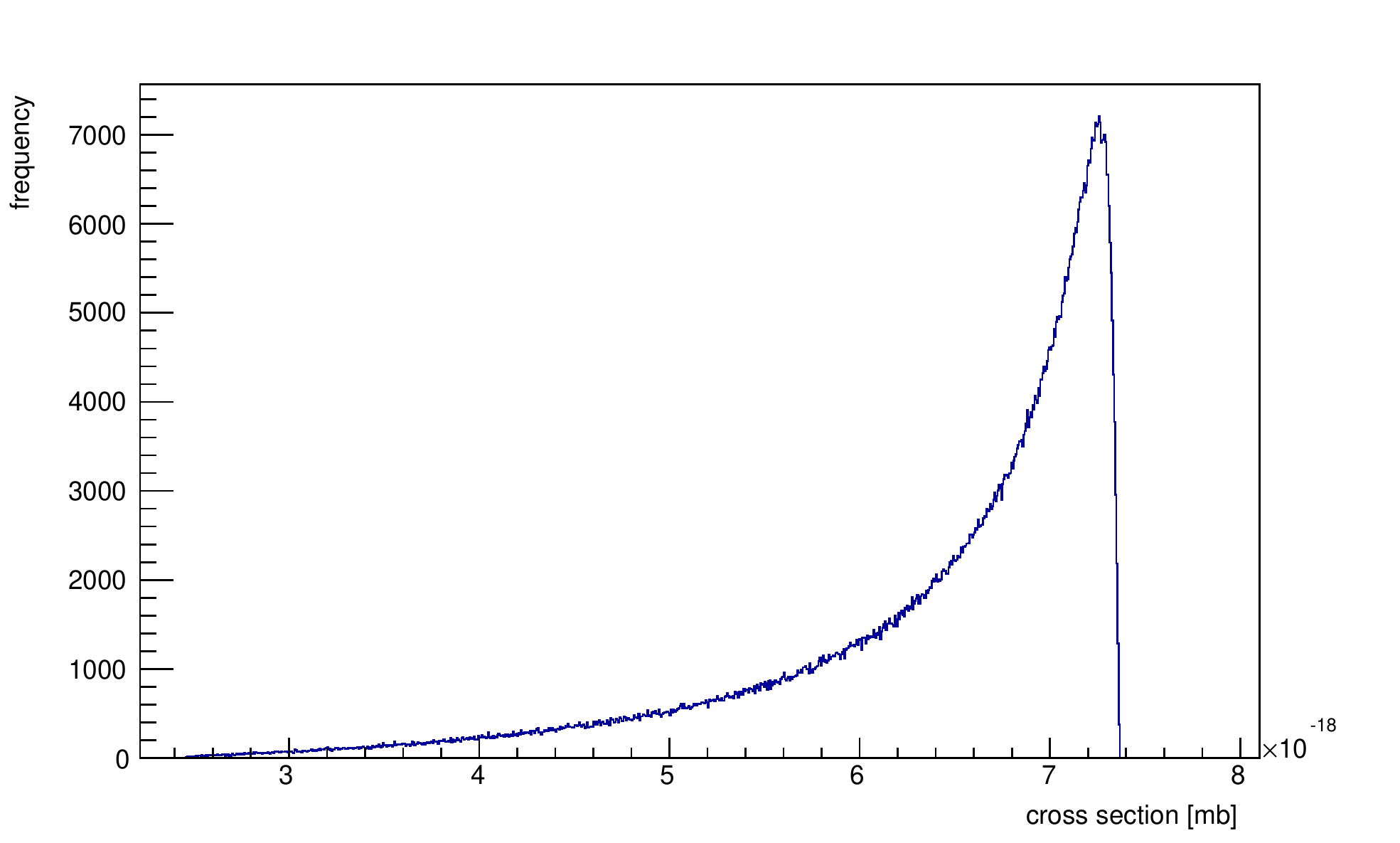}
\caption{A sample differential cross-section graph when the incident neutrino energy is $E_\nu = 0.420$~MeV.}
\label{Fig:prob_dist_res}
\end{figure}

\begin{figure}[h!tbp]
\centering
\begin{subfigure}[b]{0.45\textwidth}
	\includegraphics[width=\textwidth]{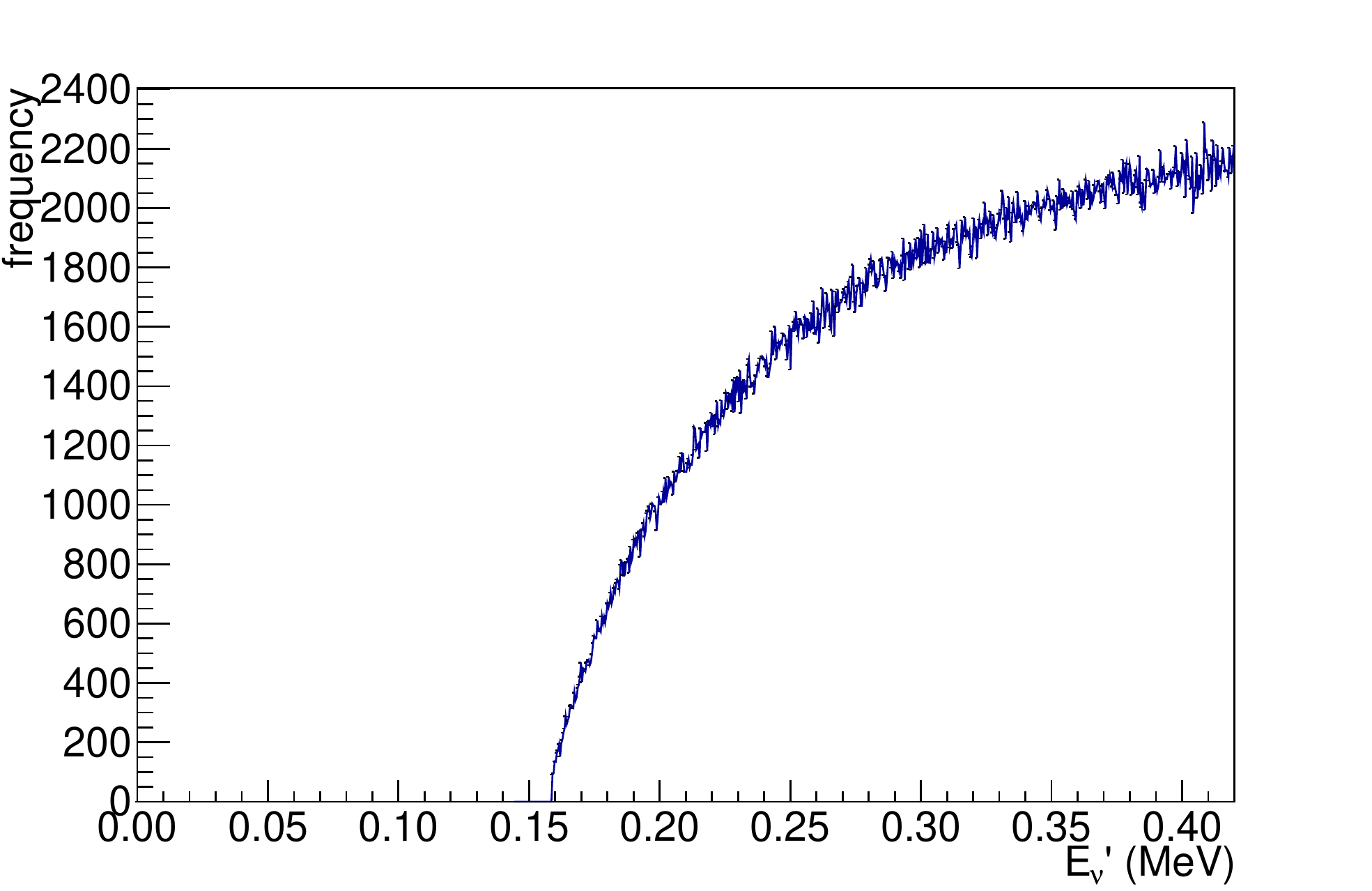}
	\caption{$E_\nu'$: the final energy of the ejected neutrino.}
	\label{Fig:Evf}
	\end{subfigure}
	~
	\begin{subfigure}[b]{0.45\textwidth}
	\includegraphics[width=\textwidth]{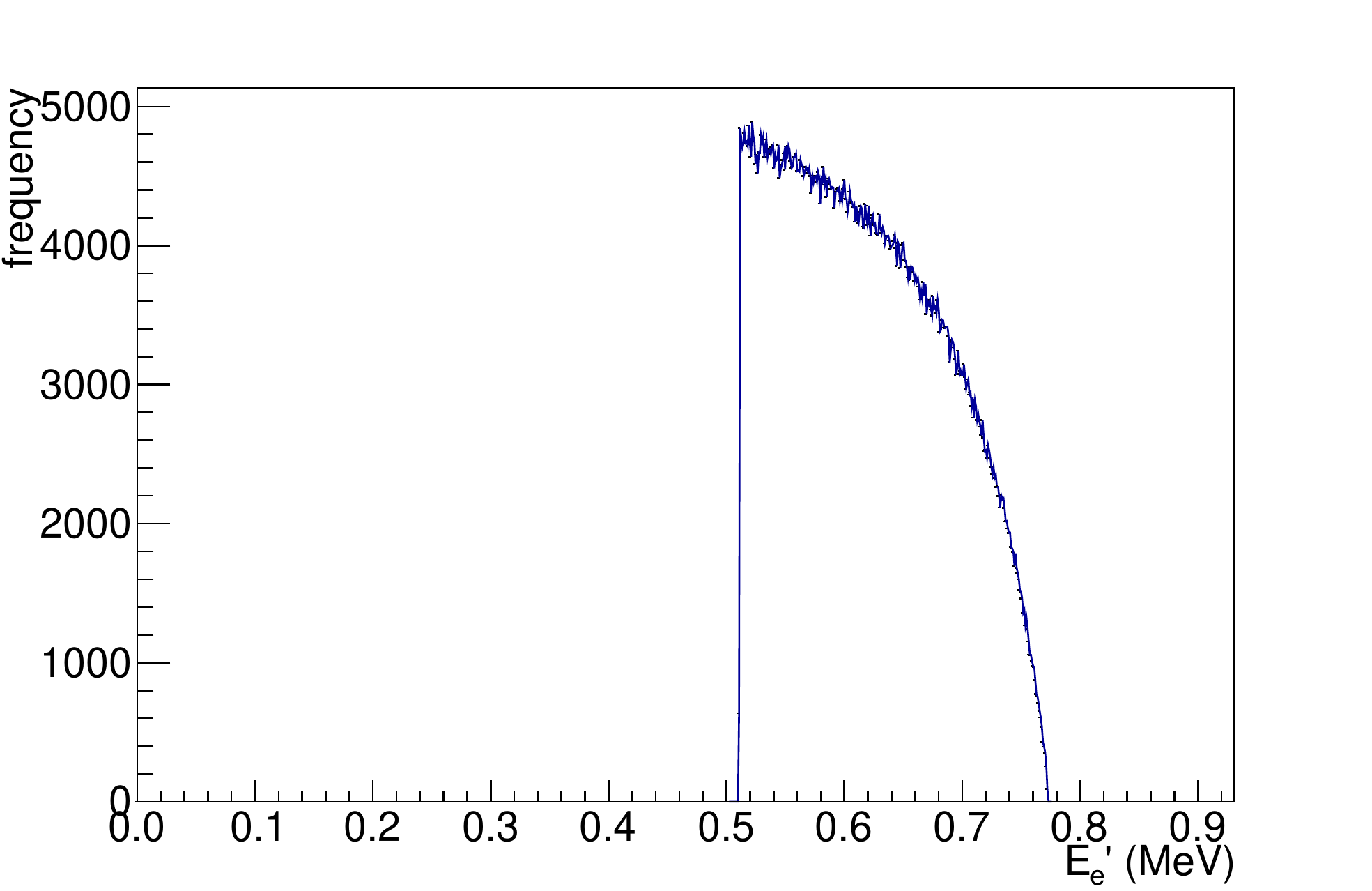}
	\caption{$E_e'$: the final energy of the ejected electron.}
	\label{Fig:Eef}
	\end{subfigure}
	
	\begin{subfigure}[b]{0.45\textwidth}
	\includegraphics[width=\textwidth]{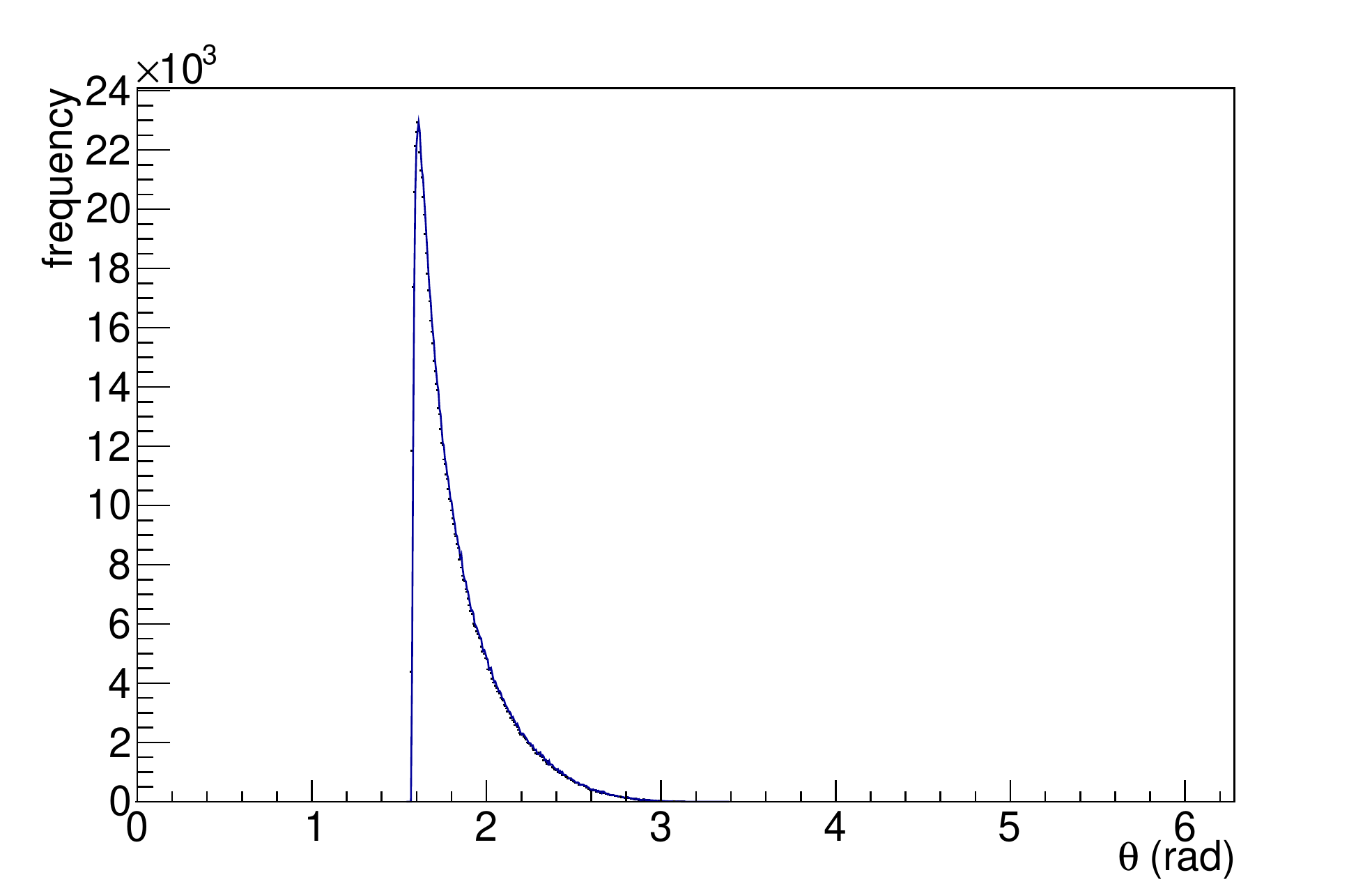}
	\caption{$\theta$: the opening angle between the electron and the neutrino.}
	\label{Fig:th}
	\end{subfigure}
	~
	\begin{subfigure}[b]{0.45\textwidth}
	\includegraphics[width=\textwidth]{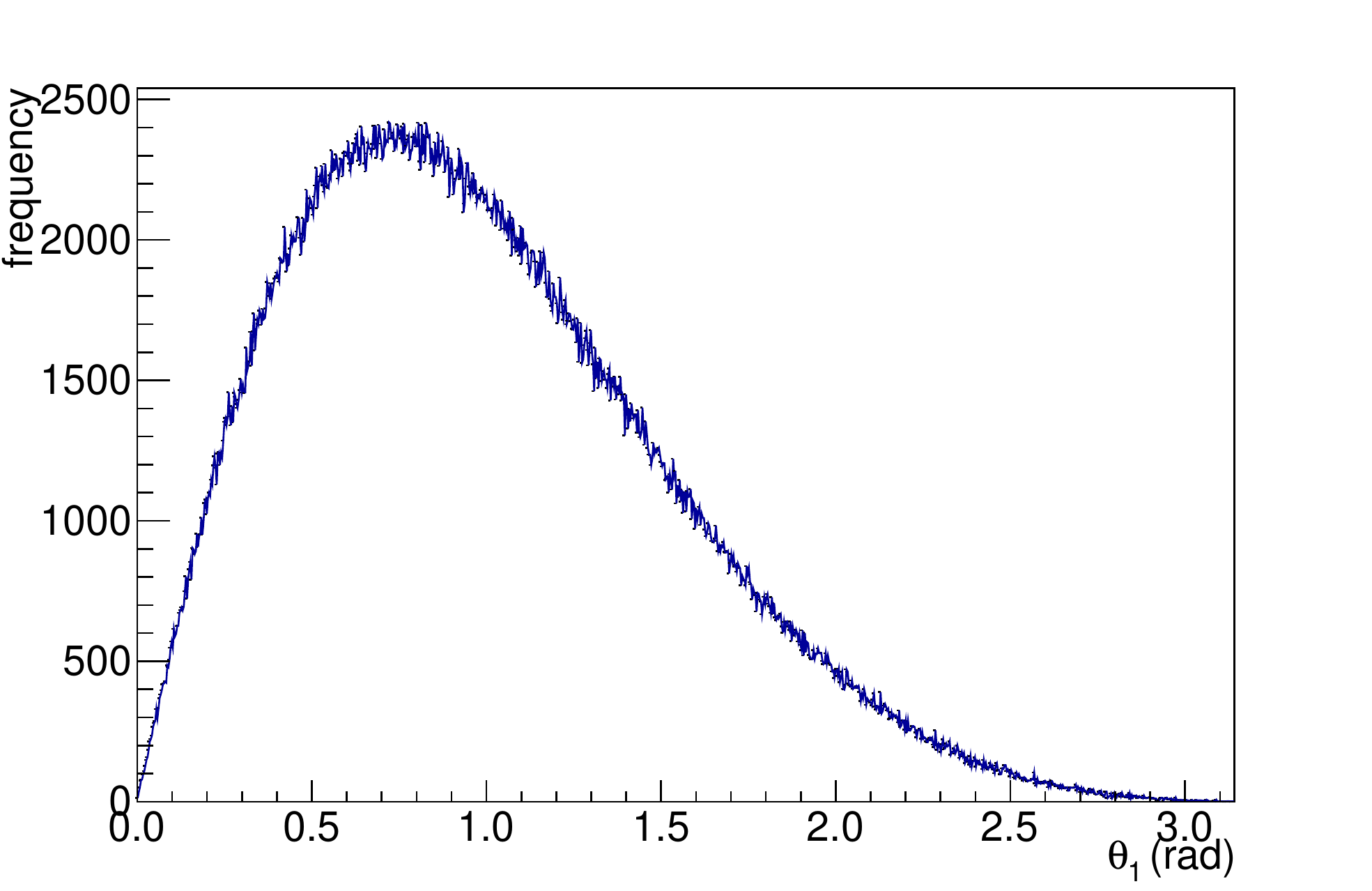}
	\caption{$\theta_1$: the angle of deflection for the neutrino from its original flight trajectory.}
	\label{Fig:th1}
	\end{subfigure}
\caption{Relevant variables that were calculated for the $E_{\nu}$ = 0.420~MeV case.}
\label{Fig:variables}
\end{figure}

In \Fig{Fig:prob_dist_res} we show the frequency for accepting the $p_{cal}$ probability values, indicating the shape of the probability density function.  
In \Fig{Fig:variables} we plot the other parameters of interest $E_e'$, $E_{\nu}'$, $\theta_1$, and $\theta$.
In \Fig{Fig:Evf} we discover that the final values for the neutrino's energy cannot be below approximately 0.16~MeV, and the neutrino is more likely to give up more of its energy.  Once the ejected neutrino's energy is known, the final energy for the electron is determined by the conservation of energy equation $E_e' = E_\nu + m_e - E_\nu'$.  The final energy for the electron can never be less than 0.511~MeV since this is the rest mass for the electron.  The negative slope of and $E_e'$ versus $E_v'$ curve implies the electron is more likely to have a lower energy.  Since the neutrino can never give up all of its energy, the electron can never receive the full 0.420~MeV of the neutrino's incident energy, which is why the $E_e'$ graph cuts off between $0.76 \text{ and } 0.78$~MeV.

We show in \Fig{Fig:th} that opening angle between the electron and the neutrino can never be less than $\frac{\pi}{2}$ because of conservation of momentum, and the distribution also peaks at~$\theta = \frac{\pi}{2}$.
In \Fig{Fig:th1} we see how this opening angle is distributed to the deflection angle of the neutrino.  The $\theta_1$ distribution is an approximately bell shaped distribution varying between 0 and $\pi$, but skewed to the right, favoring smaller deflection angles.  

\begin{figure}[h!tbp]
\centering
\includegraphics[width=\textwidth]{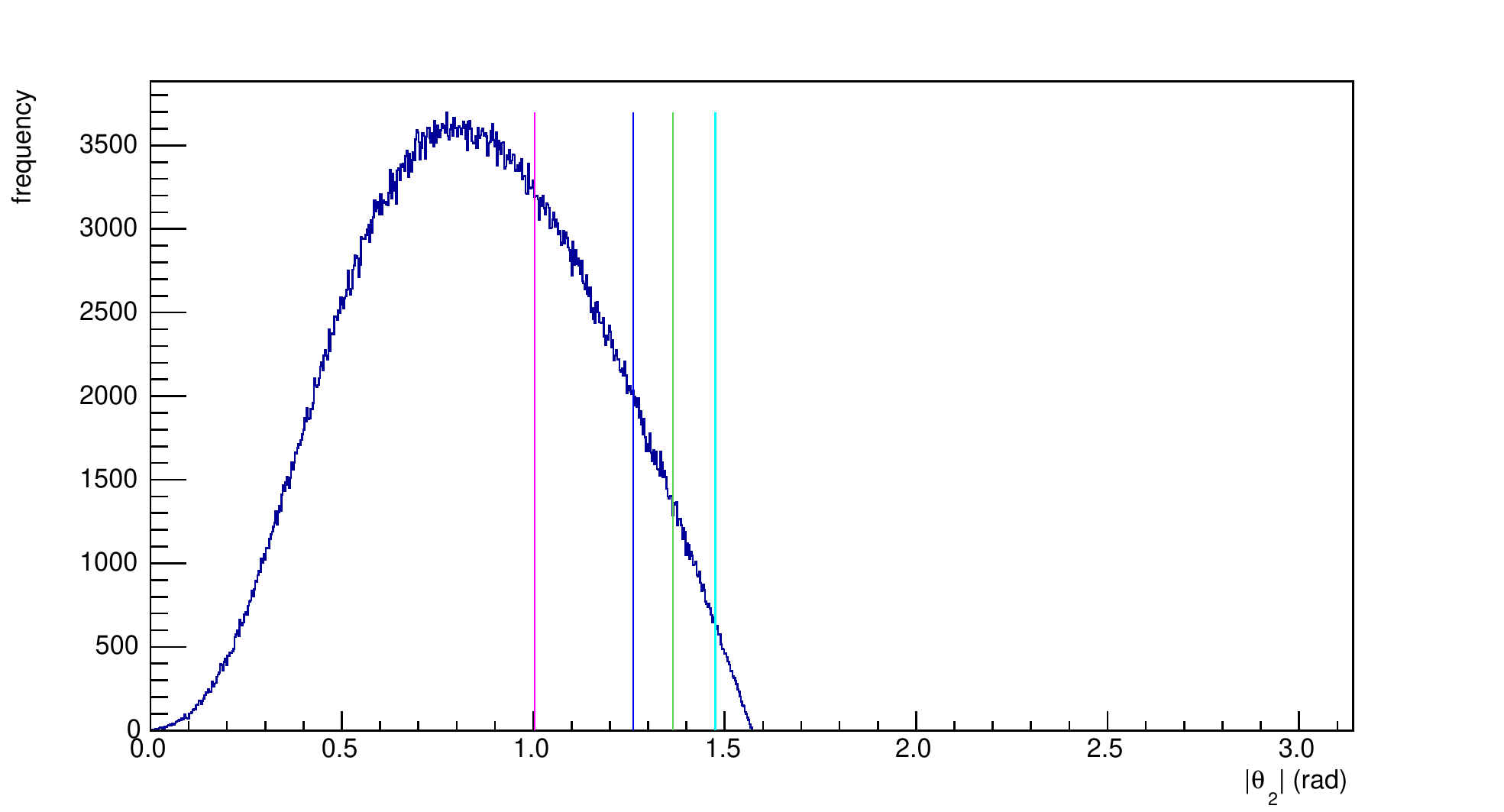}
\caption{$\theta_2$ histogram for $E_{\nu}$ = 0.420 MeV.  The vertical lines on the graph, reading from left to right, indicate the $\theta_2$ confidence limits for the 68.5\%, 90\%, 94.5\%, and 99\%, respectively.}
\label{Fig:th2_dist}
\end{figure}

Finally, we show the $\theta_2$ distribution in \Fig{Fig:th2_dist}.  Like the $\theta_1$ distribution, it is approximately bell-shaped, with a mean of 0.843 radians. 
The $1\sigma$ confidence line is shown in magenta (left-most line) on the plot, the vertical line the furthest to the left at 0.99 radians, demarcating the upper-limit for 68.5\%  of the $\theta_2$ values.  Similarly, the 90\%, 95\%, and 99\% confidence lines are shown in purple, green, and teal with the higher confidence values positioned progressively to the right.  

\begin{figure}[h!tbp]
\centering
\includegraphics[width=\textwidth]{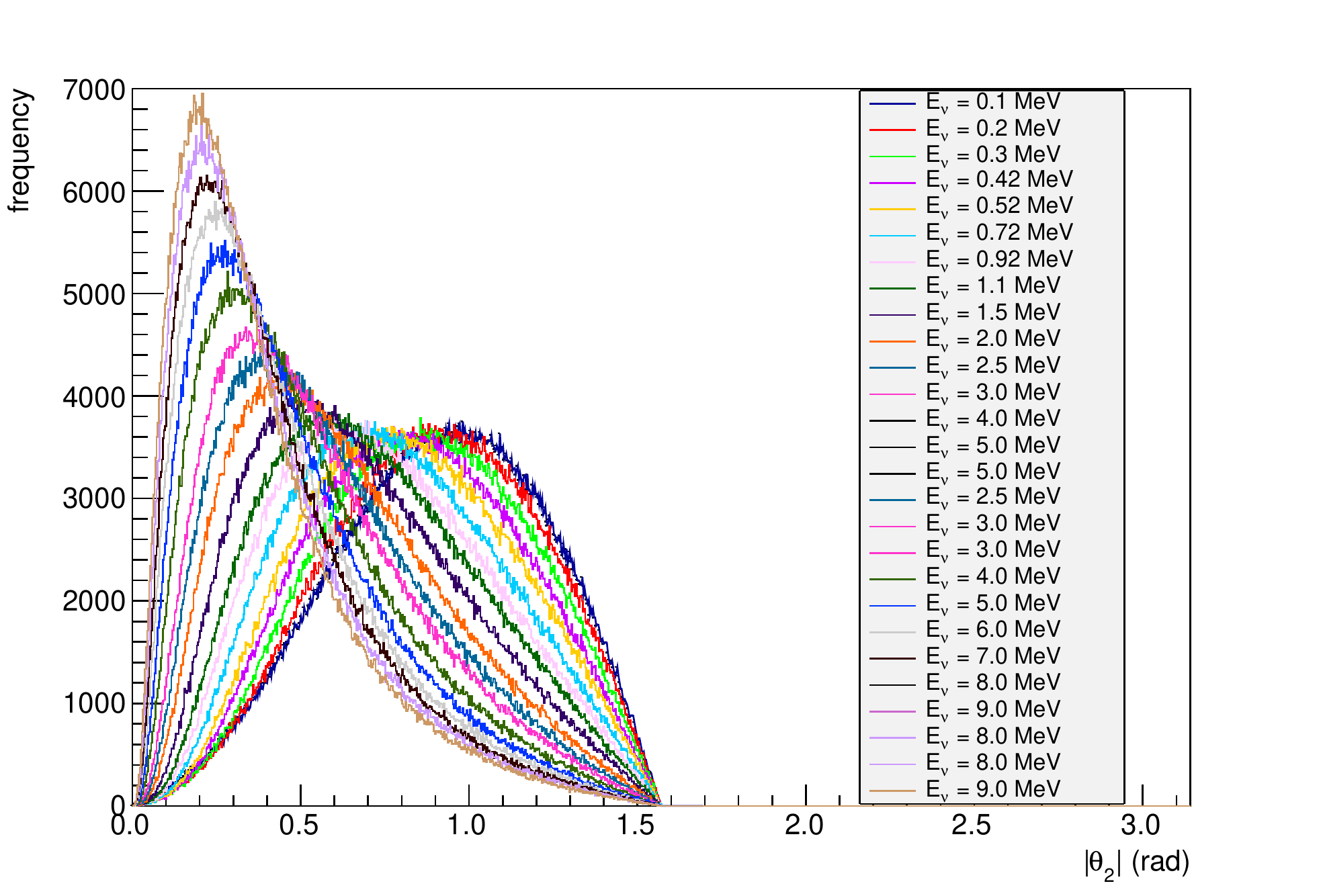}
\caption{Overlaying the histograms for allowed $\theta_2$ values for the various energies.  Increasing $E_\nu$ shifts the $\theta_2$ distribution to the left as the peak becomes more pronounced.}
\label{Fig:th2}
\end{figure}

\begin{figure}[h!tbp]
\centering
\includegraphics[width=\textwidth]{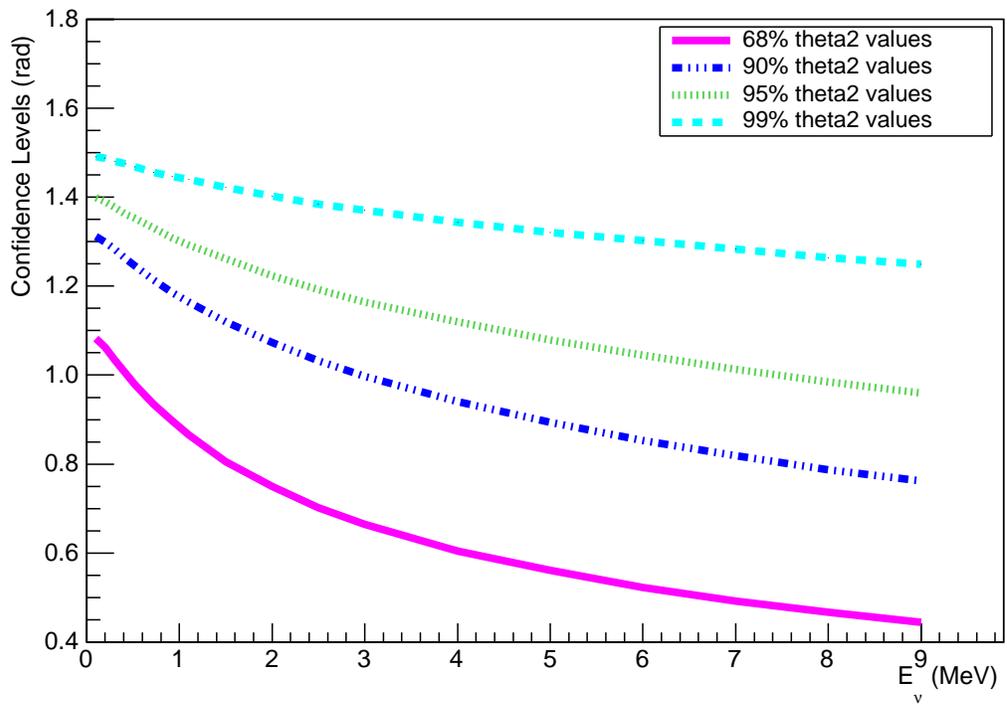}
\caption{The upper bound for $\theta_2$ for the desired confidence levels.}
\label{Fig:confidence_points}
\end{figure}

We then proceeded to find the confidence values for the $\theta_2$ histogram over a range of incident neutrino energies, $E_{\nu}$ = 0.1, 0.2, 0.3, 0.42, 0.52, 0.72, 0.92, 1.1, 2.0, 3.0, 4.0, 5.0, 6.0, 7.0, 8.0, and  9.0~MeV.  
We overlaid the $\theta_2$ histograms for each of these incident solar neutrino energies in \Fig{Fig:th2}.  The $\theta_2$ distributions become more sharply peaked for higher values for the incident neutrino energy.
For each of the $\theta_2$ plots, we calculated the 68\%, 90\%, 95\%, and 99\% confidence levels, and plotted the confidence points  as a function of energy as shown in \Fig{Fig:confidence_points}.  The 68\% confidence line is the lowest, because it does not require as many $\theta_2$ values to be below this limit.  The confidence level is higher for lower incident energy values, and plateaus for the larger incident neutrino energies.  This property was seen earlier from the overlaid $\theta_2$ plot distribution since the larger incident neutrino energies corresponded to more tightly peaked $\theta_2$ graphs.

\newpage
\newpage
\newpage

\section{Checking Results}

\subsection{Other variable dependancies for $E_\nu$ = 0.420~MeV}
\begin{figure}[h!tbp]
        \centering
        \begin{subfigure}[h!btp]{0.45\textwidth}
                	\includegraphics[width=\textwidth]{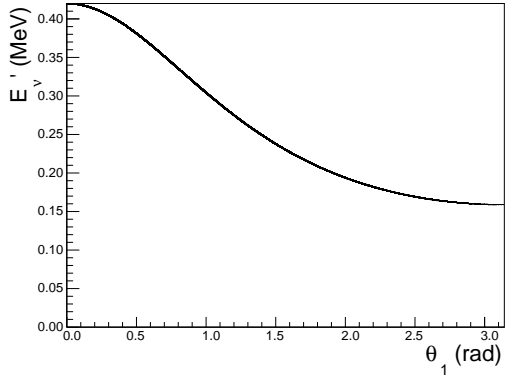}
                	\caption{Final neutrino energy versus ejected angle for the neutrino.}
                	\label{Fig:Ev_f_v_th1}
        \end{subfigure}
        ~	
        \begin{subfigure}[h!tbp]{0.45\textwidth}
            	\includegraphics[width=\textwidth]{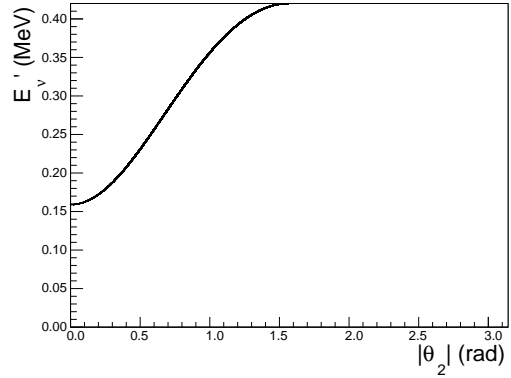}
            	\caption{Final neutrino energy versus ejected angle for the electron.}
            	\label{Fig:Ev_f_v_th2}
        \end{subfigure}
        \begin{subfigure}[h!tbp]{0.45\textwidth}
            	\includegraphics[width=\textwidth]{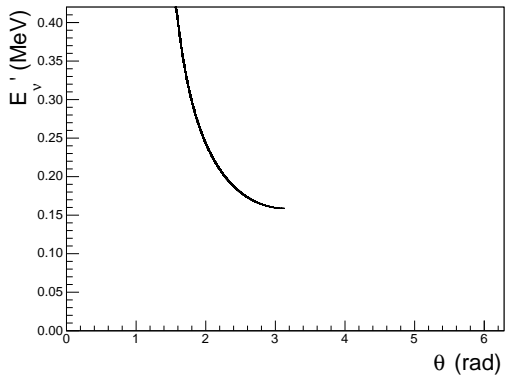}
            	\caption{Final neutrino energy versus opening angle.}
            	\label{Fig:Ev_f_v_th}
        \end{subfigure}
        \begin{subfigure}[h!tbp]{0.45\textwidth}
            	\includegraphics[width=\textwidth]{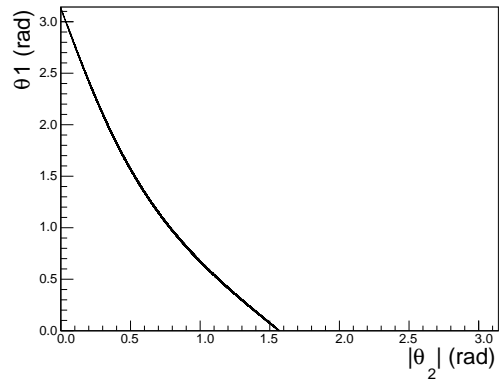}
            	\caption{Opening angle for the neutrino versus the opening angle for the electron.}
            	\label{Fig:th1_v_th2}
        \end{subfigure}
\caption{Examining how $E_\nu'$ varies as a function of $\theta_1$ and $\theta_2$.}
\end{figure}

Each of the plots depends on the calculated value of the probability distribution for the randomly generated $E_\nu'$ values.  
Since we calculate other pertinent variables after we have a physical random $E_\nu'$, we investigated the dependence of the variables with respect $E_\nu'$.
We show the dependance of the deflection angle of the neutrino with respect to the ejected energy of the neutrino in \Fig{Fig:Ev_f_v_th1}.  If the neutrino loses very little energy, it should not be expected to be deflected much from its path, and hence it should have a very narrow deflection angle.  So large $E_\nu'$ values should correspond to small $\theta_1$ values, and visa-versa, as illustrated in \Fig{Fig:Ev_f_v_th1}.

Furthermore, a larger $E_\nu'$ allows the magnitude of the opening angle for the electron to be larger while still conserving energy and momentum.  
We expect that larger $E_\nu'$ should yield a larger magnitude for $\theta_2$, which is confirmed in Figure~\ref{Fig:Ev_f_v_th2}.  
Also, $|\theta_2|$ never exceeds $\frac{\pi}{2}$, because if the electron back-scatters, momentum could not be conserved in the $\hat{z}$ direction.

The opening angle between the electron and the neutrino is defined as $\theta = \theta_1 - \theta_2 = |\theta_1| + |\theta_2|$.  
Since larger $E_v'$ tend to produce larger magnitudes for $\theta_1$, but smaller magnitudes for $\theta_2$, these two effects should compete with each other.  
It turns out that the $\theta_1$ dependence is stronger because larger $E_\nu'$ values correspond to smaller $\theta$ values according to Figure~\ref{Fig:Ev_f_v_th}.  
The $\theta$ values range between $[\frac{\pi}{2}, \pi]$.  
The largest values for $E_\nu'$ correspond to $\theta = \frac{\pi}{2}$, and this makes sense because this is the value that $\theta_2$ has when the energy is maximum.  
The smallest values for $E_\nu$ are for $\theta = \pi$, and this is when the $\theta_1$ distribution takes over, because the $E_\nu'$ is minimized when $\theta_1$ is maximized, at $\theta = \pi$.

Finally $\theta_1$ and $\theta_2$ were plotted against each other.  Since smaller $\theta_1$ values and larger $|\theta_2|$ values both corresponded to smaller $E_\nu'$ values, $\theta_1$ and $|\theta_2|$ should have a negative slope when plotted against each other.  We see this confirmation in \Fig{Fig:th1_v_th2}, which verifies that the experimental results are self-consistent.

\subsection{Increasing Incident Neutrino Energy}

As a final test, we increased the incident neutrino energy up dramatically.  For solar neutrinos, $E_\nu =$ 0.42 MeV is approximately 90\% of the electron's rest mass.  When the neutrino's energy is much larger than the electron's rest mass, we can use a simplified formula for the interaction's cross-section,\cite{Hosen}
\begin{equation}
s = (E_\nu + m_e)^2.
\label{cross_section_large_Ev}
\end{equation}

\begin{figure}[h!tbp]
\centering
\begin{subfigure}[b]{0.45\textwidth}
	\includegraphics[width=\textwidth]{prob_420keV.pdf}
	\caption{Cross section distribution for $E_\nu = 0.42$~MeV.}
	\label{Fig:prob_420keV}
	\end{subfigure}
	~
	\begin{subfigure}[b]{0.45\textwidth}
	\includegraphics[width=\textwidth]{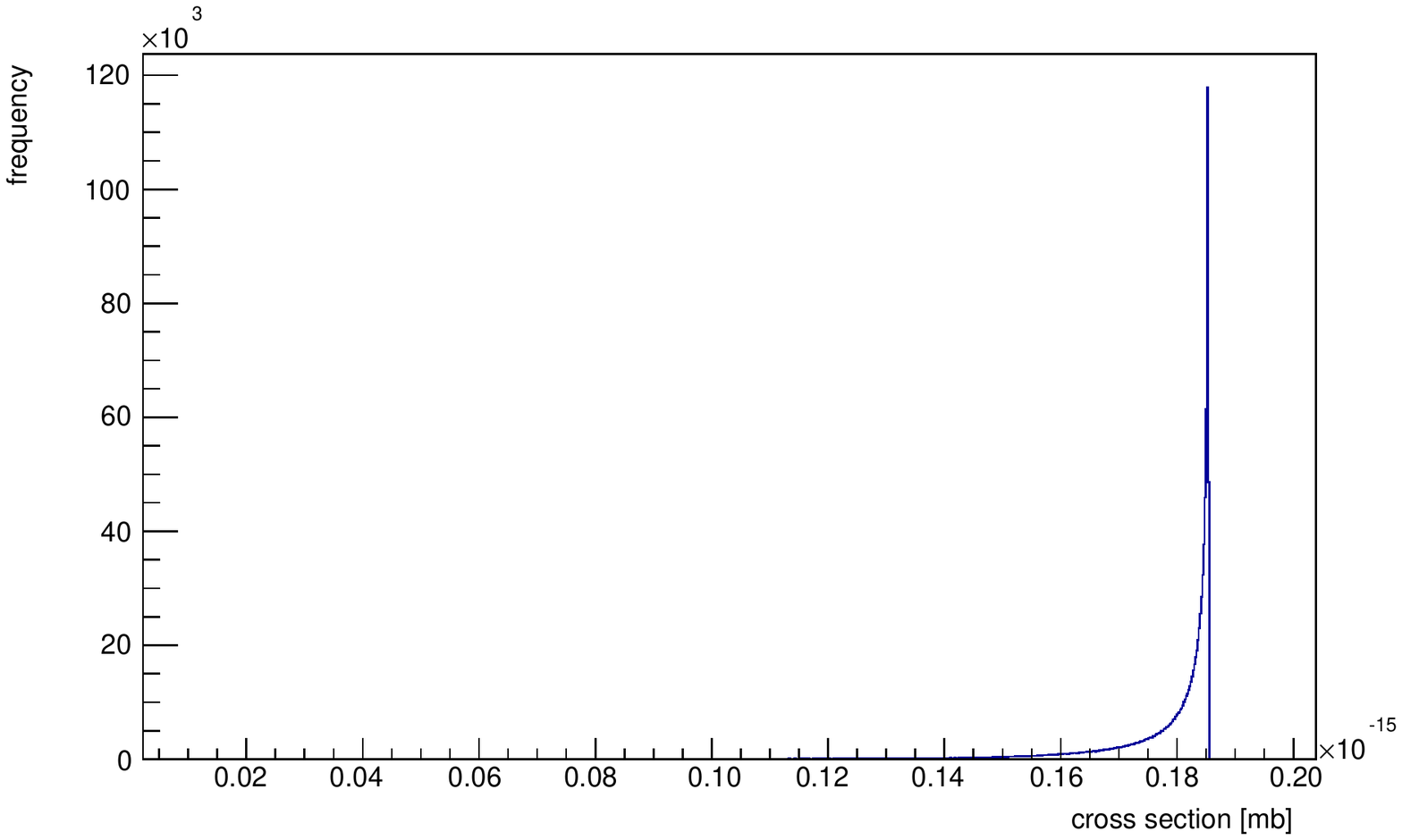}
	\caption{Cross section distribution for $E_\nu = 4$~MeV.}
	\label{Fig:prob_4MeV}
	\end{subfigure}
	
	\begin{subfigure}[b]{0.45\textwidth}
	\includegraphics[width=\textwidth]{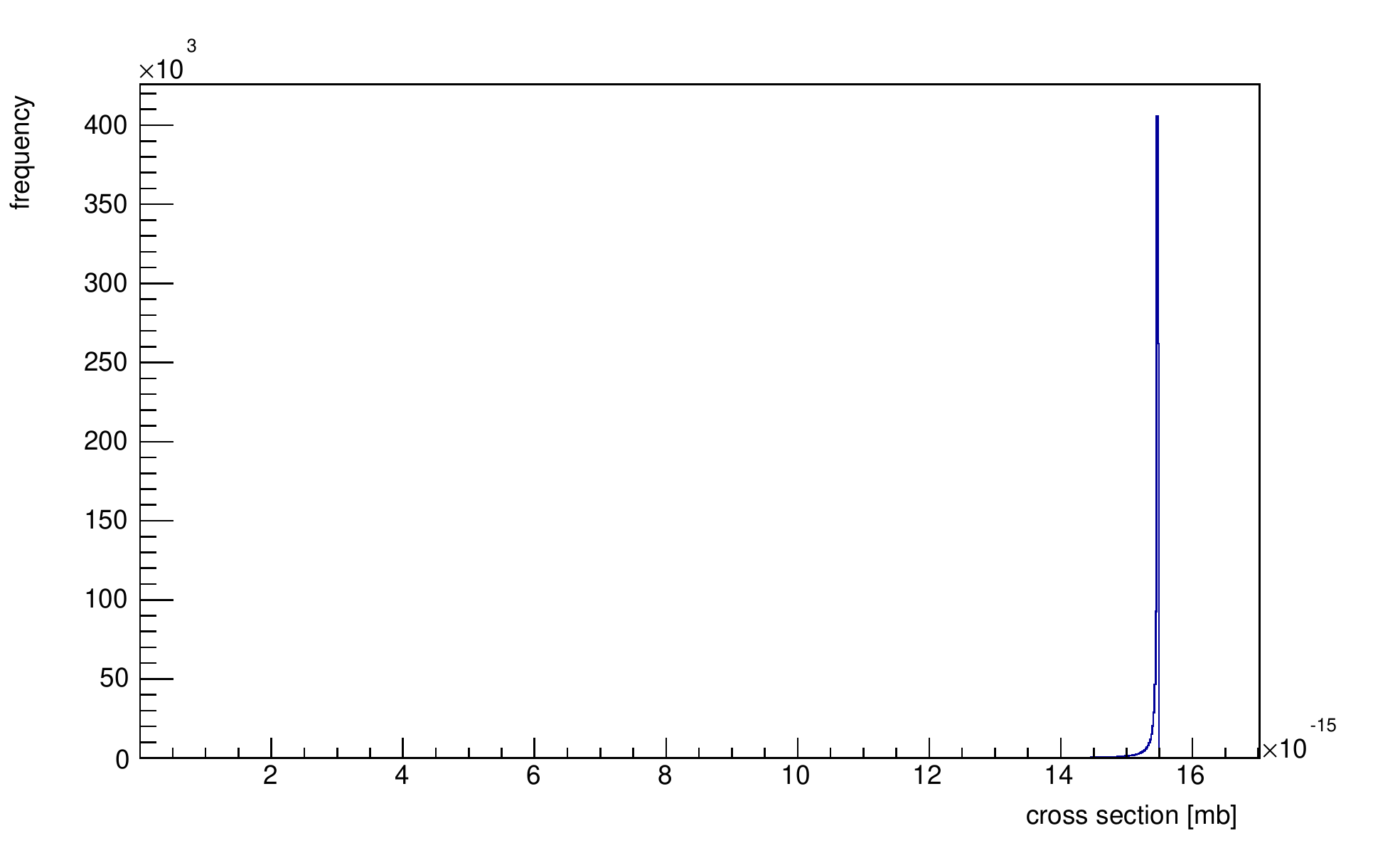}
	\caption{Cross section distribution for $E_\nu = 40$~MeV.}
	\label{Fig:prob_40MeV}
	\end{subfigure}
	~
	\begin{subfigure}[b]{0.45\textwidth}
	\includegraphics[width=\textwidth]{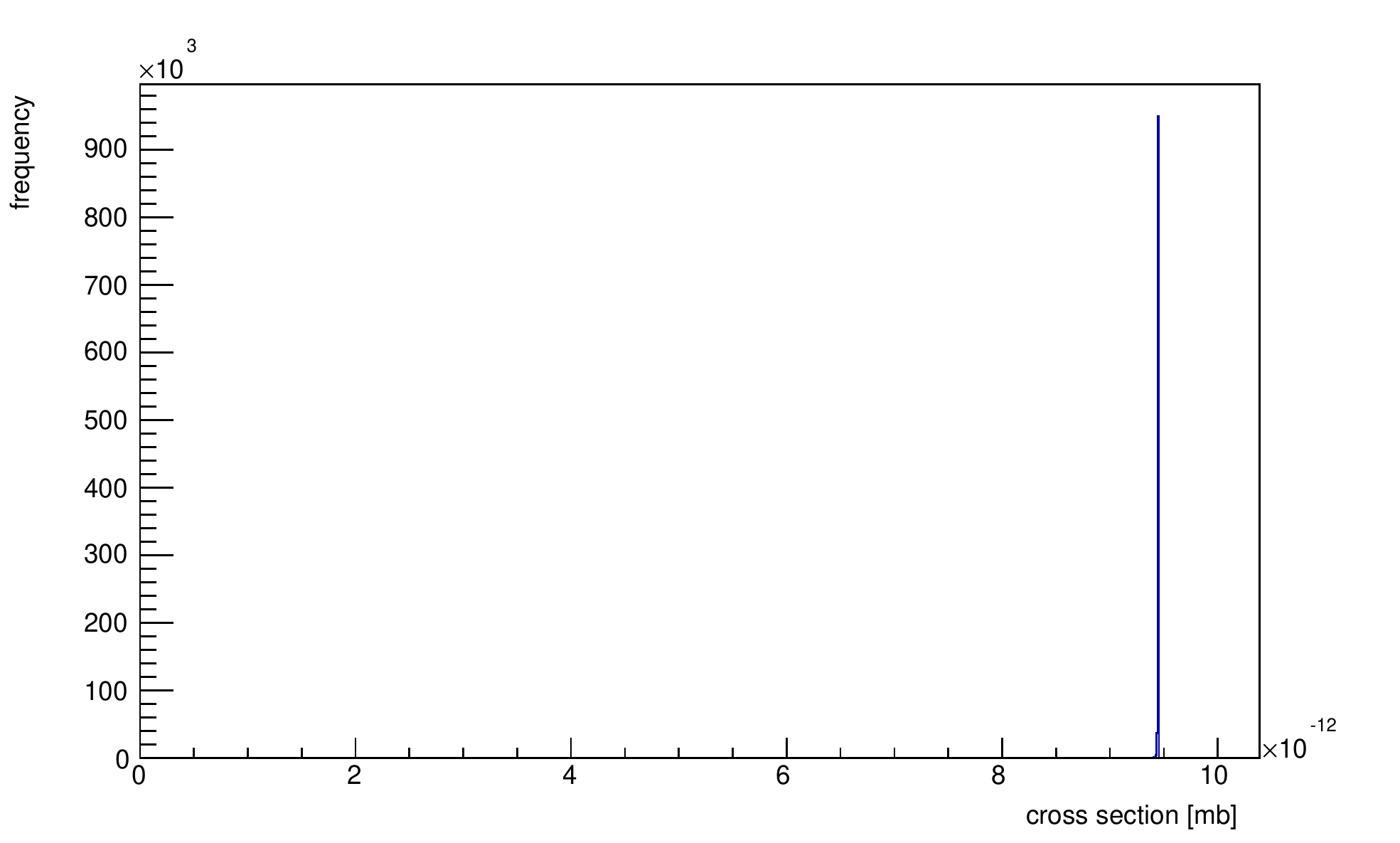}
	\caption{Cross section distribution for $E_\nu = 1$~GeV.}
	\label{Fig:prob_1GeV}
	\end{subfigure}
\caption{Plots for the probability distributions for various energies.}
\label{Fig:prob_dist}
\end{figure}

Since this constant cross section formula is valid for large enough energies, the differential cross-section becomes peaked and approaches a delta function as the incident neutrino energy is increased.  In \Fig{Fig:prob_dist} we display the differential cross-section distribution for varying incident neutrino energies $E_\nu$ = 0.42~MeV, 4~MeV, 40~MeV and 1~GeV.  
The probability distribution becomes more peaked as $E_\nu$ increases, until it is indistinguishable from a delta function when $E_\nu =$~1~GeV for the precision of the abscissa axis.
This agrees with the prediction in \Equation{cross_section_large_Ev} that the cross-section approaches a constant function for large enough incident neutrino energies.

\begin{figure}[h!tbp]
\centering
\begin{subfigure}[b]{0.45\textwidth}
	\includegraphics[width=\textwidth]{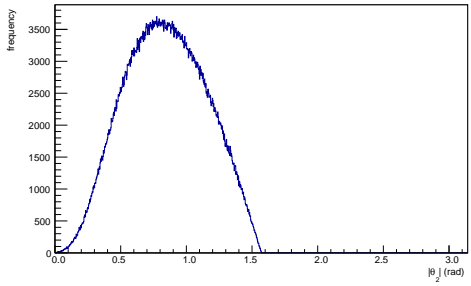}
	\caption{$\theta_2$ distribution for $E_\nu = 0.42$~MeV.}
	\label{Fig:th2_420keV}
	\end{subfigure}
	~
	\begin{subfigure}[b]{0.45\textwidth}
	\includegraphics[width=\textwidth]{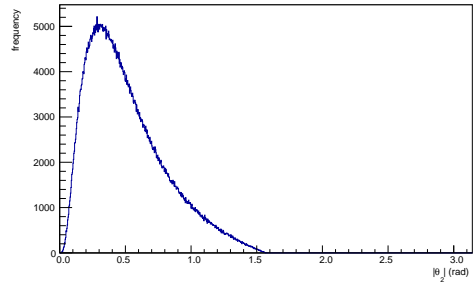}
	\caption{$\theta_2$ distribution for $E_\nu = 4$~MeV.}
	\label{Fig:th2_4MeV}
	\end{subfigure}
	
	\begin{subfigure}[b]{0.45\textwidth}
	\includegraphics[width=\textwidth]{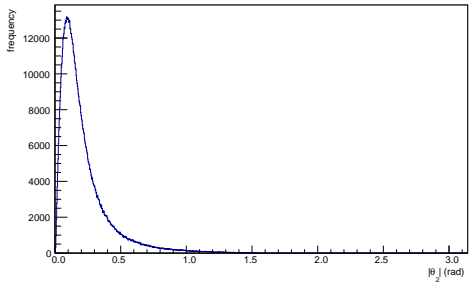}
	\caption{$\theta_2$ distribution for $E_\nu = 40$~MeV.}
	\label{Fig:th2_40MeV}
	\end{subfigure}
	~
	\begin{subfigure}[b]{0.45\textwidth}
	\includegraphics[width=\textwidth]{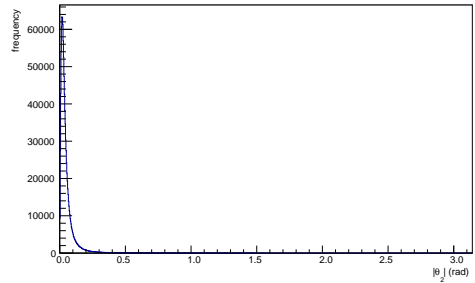}
	\caption{$\theta_2$ distribution for $E_\nu = 1$~GeV.}
	\label{Fig:th2_1GeV}
	\end{subfigure}
\caption{Plots for the $\theta_2$ distribution for various incident neutrino energies.}
\label{Fig:th2_VaryEv}
\end{figure}

As before, we use the differential cross-sections to find $\theta_2$, the ejected angle for the electron.  The more energy the neutrino has, the more energy it can impart to the electron.  More energetic electrons have smaller angles of deflection, as explained previously in the discussion for \Fig{Fig:Ev_f_v_th2}.  Therefore, larger values for the incident neutrino energy, $E_\nu$, should yield tighter opening angles for $\theta_2$, which we verify in \Fig{Fig:th2_VaryEv}.

A more peaked $\theta_2$ distribution gives us more information about the incident neutrinos flight direction, so these graphs can also explain why higher energy neutrinos from the solar neutrino spectrum shown in \Fig{Fig:sensitivities} are more reliable in assessing the neutrino flight direction in experiment.

\section{Conclusion}

Using the scattering cross-section for electron-neutrino interactions at leading order, given by relativistic quantum mechanics, we determined the probability distribution for the interaction.  From this distribution, we used the accept-reject method to find the allowed angles for the ejected electron.  This $\theta_2$ was found to be within 0.99~radians, or $57^\circ$ for the 1~sigma confidence level.  This angular resolution is sufficient to at least determine which hemisphere a given event originated in, and might therefore be used to provide a way to distinguish which ``hits'' could be due solar neutrinos in an underground dark matter experiment.  We hope this simulation encourages future development of detector technology that can provide information about the directionality of low-energy scattering events in experiments searching for rare cosmic phenomena, such as dark matter interactions.

\begin{acknowledgments}

We gratefully acknowledge R. Stroynowski for proposing this project and his helpful directions and suggestions.  We would also like to thank R. Scalise for valuable discussions throughout the course of this analysis. 
This research was supported by SMU. 

\end{acknowledgments}

\end{document}